\documentclass[prl,twocolumn,showpacs,superscriptaddress,aps]{revtex4-2}
\usepackage{graphicx,amssymb,amsmath,psfrag,color,mathtools}
\usepackage[colorlinks=true,citecolor=blue,linkcolor=blue,urlcolor=blue]{hyperref}
\begin{document}
\definecolor{darkgreen}{rgb}{0,0.5,0}
\newcommand{\be}{\begin{equation}}
\newcommand{\ee}{\end{equation}}
\newcommand{\jav}[1]{\textcolor{red}{#1}}
\newcommand{\rim}[1]{\textcolor{green}{#1}}
\newcommand{\cpm}[1]{\textcolor{blue}{#1}}

\title{Higher order OTOCs in Luttinger liquids}
\title{Signatures of chaos and $k$-design in higher-order OTOCs of Luttinger liquids}
\title{Universal saturation and $k$-design mimicry in higher order OTOCs of Luttinger liquids}
\title{Mimicry of chaos and $k$-design in higher order OTOCs of Luttinger liquids} 
\author{Bal\'azs D\'ora}
\email{dora.balazs@ttk.bme.hu}
\affiliation{Department of Theoretical Physics, Institute of Physics, Budapest University of Technology and Economics, M\H uegyetem rkp. 3., H-1111 Budapest, Hungary}
\affiliation{MTA-BME Lend\"ulet "Momentum" Open Quantum Systems Research Group, Institute of Physics, Budapest University of Technology and Economics,
M\H uegyetem rkp. 3., H-1111, Budapest, Hungary}

\author{C\u at\u alin Pa\c scu Moca}
\affiliation{Department of Theoretical Physics, Institute of Physics, Budapest University of Technology and Economics, M\H uegyetem rkp. 3., H-1111 Budapest, Hungary}
\affiliation{MTA-BME Lend\"ulet "Momentum" Open Quantum Systems Research Group, Institute of Physics, Budapest University of Technology and Economics,
M\H uegyetem rkp. 3., H-1111, Budapest, Hungary}
\affiliation{Department of Physics, University of Oradea,  410087, Oradea, Romania}

\author{Roderich Moessner}
\affiliation{Max-Planck-Institut f\"ur Physik komplexer Systeme, 01187 Dresden, Germany}

\date{\today}

\begin{abstract}
Out-of-time-order correlators (OTOCs) provide a fundamental metric for quantum chaos, but capturing the fine structure of information scrambling requires exploring their higher-order generalizations. 
Here, we systematically investigate the sequence of higher-order OTOCs in a Luttinger liquid and its lattice realization, the XXZ Heisenberg chain. 
Using bosonization and numerics, we extract the full temporal dynamics of the first three OTOCs, revealing that they 
rapidly erase memory of the initial state, 
and quickly saturate to their steady state values. 
Strikingly, we show that calculating the late time saturation values for the entire sequence of higher-order 
OTOCs maps exactly onto determining the partition function of a non-Hermitian Harper model. Through this mapping, we demonstrate that for moderately strong interactions, the steady-state OTOCs 
become parametrically small up to the seventh order, mimicking higher
$k$-design. Our results reveal that Luttinger liquids exhibit an unexpectedly
profound degree of apparent scrambling when viewed through the lens of higher-order OTOCs.

\end{abstract}

\maketitle

\paragraph{Introduction.}
Out-of-time-ordered correlators (OTOCs) have emerged as a central tool for characterizing quantum information scrambling, operator growth, 
and many-body quantum chaos\cite{kitaev,kitaev17,Garciamata,swingleprx,scienceotoc,landsman}. Originally introduced in the context of superconductivity\cite{larkin} and later connected to semiclassical chaos, OTOCs 
quantify the sensitivity of quantum dynamics to perturbations through the growth of operator noncommutativity\cite{roberts2016,Stanford2016,aleiner,garttner,junli,maldacenaprd,zhugrover,prxkeyserlingk}. In chaotic systems\cite{maldacena2016,bohrdt,roberts} this 
growth may be characterized by a Lyapunov exponent and is closely related to the spreading of quantum information across many degrees of 
freedom. 



While most previous studies have focused on the standard four-point OTOC, recent developments\cite{roberts2017,googleotoc,iaconis,fritzsch} have highlighted the importance of higher-order OTOCs 
involving larger numbers of operators. These generalized correlators probe increasingly refined aspects of operator statistics and information 
spreading, analogous to the role of higher cumulants in characterizing fluctuations beyond the variance. In chaotic systems, higher-order OTOCs 
have been proposed as sensitive diagnostics of scrambling and non-Gaussian operator growth.



Quantum fluctuations and interaction effects are particularly enhanced in one dimension, where interacting particles are often described by the Luttinger liquid (LL) picture\cite{giamarchi,nersesyan,cazalillaboson}, irrespective
of the underlying statistics of particles (fermion, boson, spin, anyon). The original excitations are replaced by bosonic collective modes, accompanied by
charge fractionalization and non-Fermi liquid behaviour.
While some of these states can be relevant for quantum technologies and quantum information processing\cite{nielsen}, the behaviour of higher order OTOCs remains largely unexplored in these systems.


We investigate the time dependence and late time value of the the first seven order OTOC using a combination of bosonization, numerics in the XXZ Heisenberg chain and by mapping the late time OTOCs to the partition function of the Harper model.
We find that the striking feature of higher-order OTOCs in a Luttinger liquid is that they can mimic several signatures normally associated with chaotic scrambling --
 ordinary and even higher OTOCs may decay or vanish at late times, their values can depend strongly on interaction strength and are mostly insensitive 
to the initial state
and higher-point functions may contain nontrivial structures beyond simple powers, like
OTOC$^{(2)}\neq$ (OTOC$^{(1)}$)$^2$.

At first sight, this resembles higher $k$-design formation\cite{roberts2017,iaconis,favaprx} and genuine many-body randomness. 
A $k$-design is a specific probability distribution that mimics the statistical properties of completely uniform random distribution.
The surprise, however, is that all of this can emerge in a completely integrable, Gaussian 
theory whose dynamics is exactly controlled by free bosonic propagators and coherent quasiparticle interference. 
The expected behavior in a LL is therefore not universal Haar-like saturation, but interaction-dependent, algebraically 
constrained late-time behavior governed by operator dimensions, exchange phases, and bosonization identities. 




\paragraph{Luttinger liquid in XXZ.}
The "conventional" out-of-time-order correlator\cite{larkin,kitaev} (OTOC$^{(1)}$) is defined as
\begin{gather}
C(t)=\langle W(t)V(0)W(t)V(0)\rangle,
\label{otoc2}
\end{gather}
where $O(t)=\exp(iHt)O_s\exp(-iHt)$ and $O_s$ is the operator in the Schr\"odinger picture. This object tells us about information scrambling in quantum systems.
Recently\cite{googleotoc}, a higher order version OTOC$^{(2)}$ was measured experimentally
as\cite{iaconis,roberts2017}
\begin{gather}
C^{(2k)}(t)=\langle \left(W(t)V(0)\right)^{2k}\rangle,
\label{otock}
\end{gather}
where the $k=1$ version reduces to the conventional OTOC, while the lowest higher order OTOC is the $k=2$ case, denoted OTOC$^{(2)}$.


Our goal is to gain some understanding of the  higher order OTOC by studying their properties in a Luttinger liquid (LL), a paradigmatic model of many-body physics which permits detailed analytical and numerical progress.
Its lattice incarnation is the spin-1/2  XXZ Heisenberg model\cite{giamarchi}, which reads 
\begin{gather}
H=J\sum_m S_m^xS_{m+1}^x+S_m^yS_{m+1}^y+\Delta S_m^zS_{m+1}^z
\label{xxz}
\end{gather}
with $J>0$. Periodic boundary conditions (PBC) are used with zero magnetization. 
For $|\Delta|<1$, the system realizes a LL with gapless low-energy excitations. The $\Delta=1$ point marks a BKT transition to an Ising antiferromagnet for  $\Delta>1$.
Within the realm of LL theory, the low energy physics of the model is described by collective bosonic modes 
\begin{gather}
H=\int \frac{dx}{2\pi} v\left(K(\pi\Pi(x))^2+\frac 1K (\partial_x\phi(x))^2\right),
\label{hamboson}
\end{gather}
where $\Pi(x)$ and $\phi(x)$ are dual fields satisfying
the regular commutation relation \cite{cazalillaboson}, 
$[\Pi(x),\phi(x')]=i\delta(x-x')$, $K=\pi/2/(\pi-\arccos\Delta)$ is the LL parameter and $v$ the effective sound velocity\cite{giamarchi,nersesyan}. The $K=1$ value corresponds to the non-interacting, $\Delta=0$ limit of Eq. \eqref{xxz}. The expectation values
are taken with respect to the ground state of $H$ for a given $\Delta$.

We focus on the OTOC of hermitian version of the vertex operator\cite{giamarchi,nersesyan,delft} 
$\cos(\lambda \phi(x))$ with
$\lambda>0$.
The $\lambda=2$ version of the vertex operator corresponds to the staggered part of the magnetization $S^z$ in the $z$ direction. The spin on the lattice also contains
a long wavelength, slowly varying component.  In order to separate these two contributions in the lattice realization, 
one can consider  $S_n^z-S_{n+1}^z$, which would be the antiferromagnetic order parameter for $\Delta>1$, defined on a bond between sites $n$ and $n+1$. 
A more symmetric version centered on site $n$ is given by the traceless operator 
\begin{gather}
V_n=S_n^z-\frac12 S_{n+1}^z-\frac12 S_{n-1}^z\approx \frac{2(-1)^x}{\pi\alpha}\cos(2\phi(x)),
\label{afm}
\end{gather}
where  $\alpha$ is the short distance cutoff, the remnant of the lattice constant in the continuum limit and $x\approx n\alpha$. Higher order corrections
from long wavelength modes are suppressed by $\alpha^2$. The local trace of $V_n^2$ over 3 sites is 3. 
In the language of quantum information theory\cite{nielsen}, this is a linear combination of three local Paulis, $Z_n/2-Z_{n+1}/4-Z_{n-1}/4$.

\paragraph{The correlation function.} 
Within the realm of LL theory, we consider the operators
\begin{gather}
W=\cos(\lambda\phi(x)),\hspace*{4mm} V=\cos(\lambda\phi(0))
\label{operators}
\end{gather} 
and will evaluate the OTOCs associated to these through Eq. \eqref{otock}. As a warm up, we start with the correlation function $\langle W(t)V(0)\rangle$, whose operator content, $W(t)V(0)$ factorizes as
 \begin{gather}
\cos(\lambda\phi_1)\cos(\lambda\phi_2)=\frac12\sum_{\sigma=\pm}e^{i\sigma c}\cos(\lambda(\phi_1+\sigma\phi_2)),
\label{linear}
\end{gather}
where $\phi_1=\phi(x,t)$ and $\phi_2=\phi(0,0)$ and we also consider spatial and temporal separation between the operators. Their commutator\cite{giamarchi} is
\begin{gather}
c=i\frac{\lambda^2}{2}\left[\phi_1,\phi_2\right]=\lambda^2\frac{K\pi}{4}\Theta(vt-|x|)
\label{commutator}
\end{gather}
with $\Theta(x)$ the Heaviside  function.
The expectation value of Eq.~\eqref{linear}, i.e. the  correlation function decays in a power law  fashion \cite{giamarchi,nersesyan,luther} in both space and time with exponent $-2K$. This arises from the $\sigma=-$ term in Eq. \eqref{linear}. In addition, it also displays a ballistic light cone at $|x|=vt$.
The power law decay turns into an exponential one at finite temperatures\cite{iucci2kf}.
This is the slowest decaying correlation function and indicates the tendency to antiferromagnetic ordering. 
Indeed, this is exactly what happens for $\Delta>1$ in the XXZ Heisenberg model.

\paragraph{OTOC$^{(1)}$.}
The OTOC$^{(1)}$ behaves  qualitatively differently. 
The square of the operators in Eq. \eqref{linear} is evaluated as  
\begin{gather}
\left(\cos(\lambda\phi_1)\cos(\lambda\phi_2)\right)^2=\nonumber\\
=\frac{\cos(2c)}{4}\left(1+\cos(2\lambda\phi_1)+\cos(2\lambda\phi_2)\right)+\nonumber\\
+
\sum_{\sigma=\pm}\frac{e^{2i\sigma c}}{8}\cos(2\lambda(\phi_1+\sigma\phi_2)).
\label{square}
\end{gather}
Its expectation value in the late time limit does not decay to zero but approaches a finite value for $(vt,|x|)\gg\alpha$ from the constant term on the r.h.s.:
\begin{gather}
C^{(2)}=\langle\left(\cos(\lambda\phi_1)\cos(\lambda\phi_2)\right)^2\rangle\longrightarrow\frac 14\cos\left(2c\right),
\label{c2}
\end{gather}
plus some additional spatio-temporal power law  decaying terms from the $\sigma=-$ contribution with very fast decay due to the exponent $-8K$, revealing the light cone, 
similarly to $c$ in Eq.~\eqref{commutator}.
 Here, the steady state value of Eq.~\eqref{c2} is actually independent of the initial state and is expected to be the same 
for finite temperatures. 
By contrast, the additional decaying terms are affected by the state of the system.

For $\lambda=2$, the asymptotic value is $\frac14 \cos(2K\pi)$ inside the light cone, $vt\gg |x|$, while it approaches $\frac 14$ outside of the light cone. 
In the non-interacting, $K=1$ limit, these two values coincide while
in the presence of interaction, there is a  jump upon crossing the light cone. Also its value is interaction dependent inside the light cone only.
The most notable feature is the vanishing of $C^{(2)}$ for $K=3/4$, corresponding to $\Delta=0.5$, indicating fast scrambling in a LL and generating unitary 2-design for the dynamics. 

Unitary $k$-designs\cite{roberts2017} are random unitaries simulating up to the $k$th order statistical moments of uniformly distributed random unitaries (Haar measure). Since $C^{(2)}$ contains two forward and to backward time evolution operators in the 2 $W(t)$ operators in Eq. \eqref{otoc2}, it is structurally a 2nd moment polynomial of the unitary ensemble. A unitary 2-design formation will yield the exact same average 1st order  OTOC value as in a truly random quantum system. In a similar vein, a vanishing $k$th order OTOC would imply unitary $k$-design formation.


\begin{figure}[t!]
\centering  
\includegraphics[width=7cm]{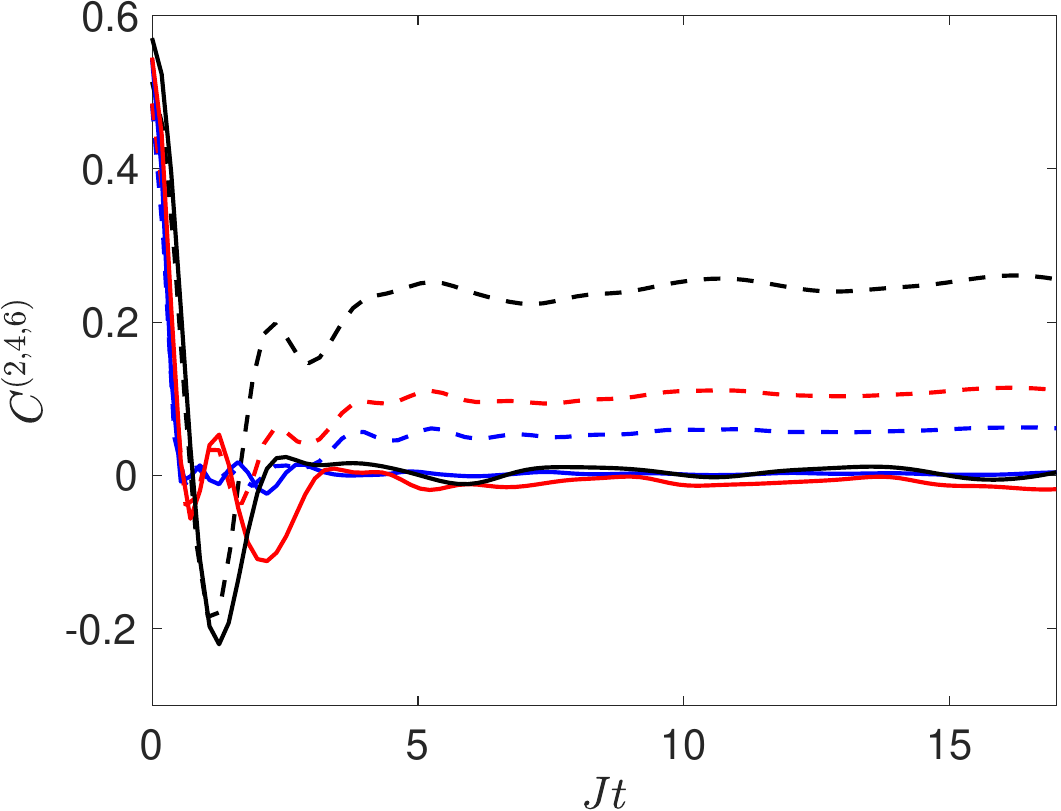}
\caption{The time evolution of the $C^{(6)}$ (blue), $C^{(4)}$ (red) and 
$C^{(2)}$ (black)
of Eq. \eqref{afm} is shown for the XXZ model with $N=26$
for $\Delta=0.2$ (dashed lint) and 0.6 (solid line).
After an initial transient time $\sim 1/J$, the OTOCs saturate fast to their steady state values.
For $\Delta=0.6$ (solid lines), the steady state values are almost vanishing and mimic high scrambling with unitary 2-, 4- and 6-design structures.
}
\label{xxzotocstime}
\end{figure}

\paragraph{OTOC$^{(2)}$}
For the higher order OTOC, one could first consider OTOC$^{(1.5)}$,  $C^{(3)}=\langle \left(W(t)V(0)\right)^3\rangle$. However, this vanishes in the long time limit, similarly to the vanishing correlation function $\langle W(t)V(0)\rangle$. 
The 
interesting object is the second order OTOC, the fourth power of the operator in Eq. \eqref{linear}, and we use the square of Eq.~\eqref{square}  to find that in the late time limit,
\begin{eqnarray} 
C^{(4)}&=&\langle\left(\cos(\lambda\phi_1)\cos(\lambda\phi_2)\right)^4\rangle\nonumber\\ &\longrightarrow&\frac{2}{16}\cos^2\left(2c\right)
+\frac{1}{64}\cos(4c), 
\label{c4} 
\end{eqnarray} 
plus other power law decaying terms as before. The square of each term in Eq.~\eqref{square} contributes to Eq.~\eqref{c4} and their expectation
values are
independent from the initial states.
Here, the first term on 
the r.h.s.\ is twice the square of the conventional OTOC, thus 
\begin{gather}
C^{(4)}=2\left(C^{(2)}\right)^2+\frac{1}{64}\cos(4c), \hspace*{3mm} (vt,|x|) \gg\alpha,
\end{gather}
where power law decaying terms are neglected again.
The diagonal contribution\cite{googleotoc} to, or disconnected part of, $C^{(4)}$ is $(C^{(2)})^2$. After  taking the expectation
value of the square of Eq.~\eqref{square}, the $\cos(2c)/4$ term  on the r.h.s. gives the diagonal contribution.
In addition, the off-diagonal or connected contributions to OTOC$^{(2)}$ arise as well, producing the remaining
$\left(C^{(2)}\right)^2+\cos(4c)/64$ terms, which require all 4 factors in $(\cos(\lambda\phi_1)\cos(\lambda\phi_2))^4$ to interfere constructively such that their expectation value is finite at late times.
 The connected and disconnected contributions are in general comparable in magnitude. However,  the  connected/disconnected  contributions in the late time limit can be made to disappear separately by fine tuning the LL parameter $K$ through interactions.

 The asymptotic value of $C^{(4)}$ is independent of the actual state of the system.
In the non-interacting limit, its value is $\left({3}/{8}\right)^2$, irrespective of $x$ and $t$. 
The very same value is taken for the interacting system outside of the light cone. As opposed to that, its value in the interacting case 
is reduced to 
$\frac{5}{32}\cos^2(2K\pi)-\frac{1}{64}$ inside the light cone. It can also take negative values for $0.38\lesssim\Delta\lesssim0.63$. The sign change occurs when
 the LL parameter satisfies
$\cos(2K\pi)=\pm 1/\sqrt{10}$ and  originates from the non-hermitian nature of the basic building
block of this OTOC structure, $W(t)V(0)$.

\paragraph{OTOC$^{(3)}$}
Evaluating the third-order OTOC, $2k=6$, yields the steady-state value
\begin{gather}
C^{(6)}=\langle\left(\cos(\lambda\phi_1)\cos(\lambda\phi_2)\right)^6\rangle
\nonumber\\ \longrightarrow   \frac{66\cos(2c)+31\cos(6c)+3\cos(10c)}{1024}
\label{c6}
\end{gather}
in the $(vt,|x|) \gg\alpha $ limit.  This  vanishes at $K=3/4$, similarly to OTOC$^{(1)}$ and forms a 6-design
\footnote{We use $\cos(6c)=4\cos^3(2c)-3\cos(2c)$ to rewrite  all harmonics but the highest (i.e. $10c$) in terms
of $\cos(2c)$ when comparing to numerics.}, and its non-interacting, $K=1$ value is $(5/16)^2$.
 Note the $10c$ high "frequency" term 
in $C^{(6)}$, which  
originates from separating the $\phi_1$ and $\phi_2$ terms from each other in the operator product. 
By taking a naive 6th power of Eq. \eqref{linear}, one would end up with $6c$ frequency only. However, since $\phi_{1,2}$ do not commute,  an additional $4c$ frequency is generated from their commutators.

\paragraph{Relation to the Harper model.}
Calculating the steady state or the time independent part of the higher order OTOC of the vertex operator in a LL
maps exactly to the partition function of the Harper model\cite{harper,aubryandre,quillen,moghaddan}. The mapping works only for the steady state but not for the full time dependence of OTOC.

The basic building block of our calculation is the operator 
$\left(\cos(\lambda\phi_1)\cos(\lambda\phi_2)\right)^{2k}$. It is expanded as
$\sum_{n_1,n_2,n_3,n_4=0}^{2k}f(c,\{n_j\})
\exp(i(n_1-n_2)\lambda\phi_1)\exp(i(n_3-n_4)\lambda\phi_2)$, where $\sum_{j=1}^4 n_j=2k$ and $f$ is a combinatorial factor, which depends also on the commutator $c$ of the $\phi_{1,2}$ fields. Among the terms in the sum, the steady state is determined
by those with no $\phi_1$ and $\phi_2$ dependence, namely with $n_1=n_2$ and $n_3=n_4$. 
We also note that due to this, the steady state of  $2k+1$ powers vanishes, as observed above.

Eq. \eqref{linear} can be rewritten as
\begin{gather}
\cos(\lambda\phi_1)\cos(\lambda\phi_2)=M\cos(X)+\bar M\cos(P)\equiv H_h
\label{harperham}
\end{gather}
with $X=\lambda(\phi_1+\phi_2)$ and $P=\lambda(\phi_1-\phi_2)$, and $M=e^{ic}/2$ is a complex number. 
Their commutation relation is that of generalized position and momentum, i.e. $[X,P]=i4c$. The $H_h$ operator
is a non-hermitian variant of Harper Hamiltonian\cite{harper,moghaddan}, where $c$ plays the role of the magnetic flux. Since the value of $c$ from Eq. \eqref{commutator} depends sensitively on the spatio-temporal separation of $\phi_{1,2}$, so do the properties of $H_h$.

Inside the light cone, $c=\lambda^2 K\pi/4$ and the problem is quantum mechanical since $X$ and $P$ do not commute with each other. Outside the light cone, $c=0$ from Eq. \eqref{commutator}, $X$ and $P$ commute and behave as classical variables and the classical limit of $H_h$ must be considered. Crossing the light cone  then also means a quantum to classical transition in interpreting  the late time value of the OTOC in terms of the Harper model. 

We focus on the steady state OTOC inside the light cone and treat $X$ and $P$ quantum mechanically as non-commuting variables. The $H_h$ Hamiltonian in Eq. \eqref{harperham} becomes hermitian for  $c=\pi\times$ integer. This requires $\lambda^2 K/4$ to be an integer, which, for $\lambda=2$, is realized in the non-interacting limit of the LL Hamiltonian in Eq. \eqref{hamboson} with $K=1$. 
Away from the non-interacting limit, the Harper
 model in Eq. \eqref{harperham} is non-hermitian in general and is self dual under the quantum Fourier transform, which maps $X\rightarrow P$ and $P\rightarrow -X$, resulting in $H_h\rightarrow H_h^+$.
Consequently, the energy spectrum is perfectly symmetric across the real axis in the complex plane and consists of complex conjugate pairs.
We note in passing that the Harper model can also be formulated in terms of the deformed Weyl algebra\cite{xuortiz} with the deformation parameter $\exp(2ic)$

The late time value of the higher order OTOC amounts to calculating the moments of the operator $H_h$, which is then equivalent to calculating the partition function
of the Harper model as
\begin{gather}
Z=\textmd{Tr}\left(\exp(-\beta H_h)\right)=\sum_{n=0}^n \frac{(-\beta)^n}{n!}\textmd{Tr}((H_h)^n),
\end{gather}
where $\beta$ is the dimensionless inverse temperature and $C^{(2k)}=\textmd{Tr}((H_h)^{2k})/d_h$ with $d_h$ the dimension of the Hilbert space of the Harper model,
and $\textmd{Tr}((H_h)^{2k+1})=0$. All of its moments are real due to the self duality of the non-hermitian Harper Hamiltonian.
We have checked numerically that by using the matrix representation\cite{korsch} $X$ and $P$ for the Harper model, its moments reproduce Eq. \eqref{c2}, \eqref{c4} and \eqref{c6}.

\begin{figure}[h!]
\centering
\includegraphics[width=8cm]{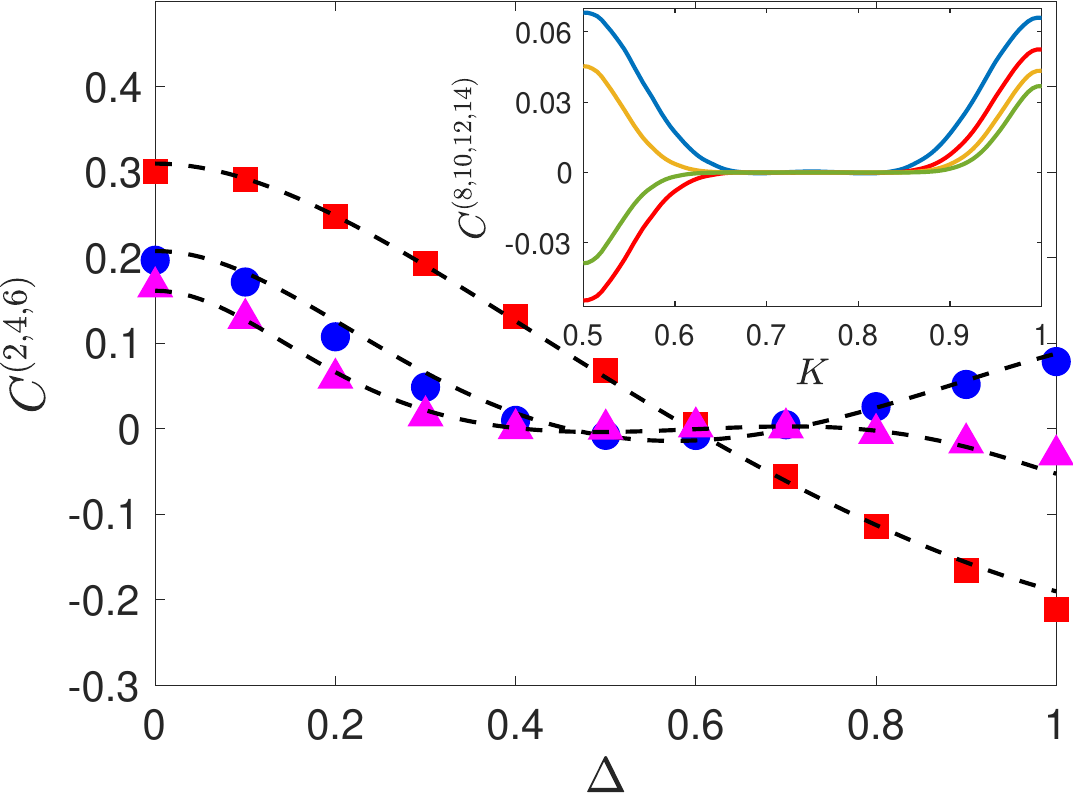}
\caption{The long time limit of the $k=1$ (red square), 2 (blue circle) and 3 (magenta triangle) OTOC of the staggered part of the spin density is 
shown for the XXZ chain with $N=26$ inside the light cone.
The symbols are the numerical data from ED, the lines are Eqs. \eqref{c2}, \eqref{c4} and  \eqref{c6} with $\lambda=2$ and $c=K\pi$. The $\cos(2c)$ is shifted by 0.24 for the $k=1$, 2 OTOCs and by 0.16 for the $k=3$ version
to account for the background signal in the numerics due to
 a) additional power law decaying terms, b) the long wavelength part of the spin density and c) interplay between long and short wavelength terms. No other fitting parameters are used.
 The inset shows the $k=4$, 5, 6, 7 OTOC in the long time limit for the LL from top to bottom at $K=1$, calculated from the mapping to the Harper model.}
\label{xxzotocs}
\end{figure}

\paragraph{Numerics.}
These analytical results for $\lambda=2$ are compared with many-body exact diagonalization of the XXZ Heisenberg chain by focusing on the local OTOC of Eq. \eqref{afm} 
in Figs. \ref{xxzotocstime} and \ref{xxzotocs}
for the representative time evolutions and steady state values. The operator of interest is $V_2$, but the results do not depend on the actual lattice site due to
PBC. We have also taken the expectation values in OTOC with respect to the ground state of the $\Delta=0$ system but time evolved the operators  with the interacting, 
$\Delta>0$ Hamiltonian. They agree very well (not shown) with the data in Fig. \ref{xxzotocstime} and confirm the bosonization prediction of relative initial state
independence of  $C^{(2,4,6)}$.
The steady state OTOCs are evaluated from averaging over $7/J$ time window before finite size effects appear. 
From Eq. \eqref{afm}, the short distance cutoff $\alpha$, which plays the role
of correlation amplitude,  can also acquire some mild $\Delta$ 
dependence\cite{lukyanov}, in addition to the additional power law terms from the $\sigma=-$ contribution in Eq. \eqref{square}.
Other operators\cite{doraotoc,motrunich,Gisti2025} such as $S_n^z$ or $S_n^z-S_{n+1}^z$ display very similar dynamics albeit the long wavelength part of the spin density is more dominantly
present in their time evolution.

In the inset of Fig. \ref{xxzotocs}, the $k=4-7$ order OTOC is plotted for the LL as a function of the LL parameter $K$ in the long time limit, calculated from the mapping to the Harper model. The $k=1,2,3$ OTOCs agree with those in the main panel. The common feature is the vanishing of these high order OTOCs over an extended region in $K$, i.e. LL parameter space. 

\paragraph{Discussion.}
Remarkably, the OTOCs  do distinguish  crisply the difference between a non-interacting and an interacting LL: for the former, simple factorization of the expectation values holds, while the latter exhibits the mimicry of chaos and $k$-design described above.
For non-scrambling dynamics, the late time limit of OTOC is expected\cite{hosur,roberts2017} to factorize as $C^{(2k)}\rightarrow \langle W^{2k}\rangle \langle V^{2k}\rangle$. For our choice of operators in Eq. \eqref{operators}, $\langle W^{2k}\rangle= \langle V^{2k}\rangle=\binom{2k}{k}/2^{2k}$. These agree with Eqs. \eqref{c2}, \eqref{c4} and \eqref{c6} in the non-interacting case, $K=1$,  and indicate that these are perfect non-scramblers from an OTOC perspective, as expected. The surprise is that  away from the non-interacting limit, the above factorization breaks down and the OTOCs mimic chaotic information scrambling, especially at around $K=3/4$.


\paragraph{Conclusions.}

Inspecting the OTOC of the staggered magnetization in LLs, we find that complex higher order correlations in OTOC do not necessarily imply chaos as many fingerprints of chaotic behaviour can be observed in LLs despite their integrability.
We find that the  OTOCs do not display simple factorization of expectation values, the late time
value of all OTOCs is relatively independent from the initial state and 
can be parametrically small for moderately strong interaction, pointing towards operator spreading. The agreement between bosonization and numerics on the XXZ model is remarkable for the saturation value of the first three OTOCs.

We also demonstrate that the problem of calculating the late  time value of all higher order OTOCs in LL is mapped to  the partition function of a non-hermitian variant of the Harper model. As a proof of principle, we use this to obtain the first 7 order OTOCs for the LL, which can be made vanishingly small for moderately strong interactions.
All in all, LLs can look highly scrambled from an OTOC perspective as  OTOCs detect operator growth, but not necessarily operator complexity.

\begin{acknowledgments}
We are grateful to Felix Fritzsch, Sarang Gopalakrishnan, Bal\'azs Het\'enyi and Philippe Suchsland for discussions.
This work was supported by the National Research, Development and Innovation Office - NKFIH  Project No. K142179,
by a grant of the Ministry of Research, Innovation and
 Digitization, CNCS/CCCDI-UEFISCDI, under projects number
PN-IV-P1-PCE-2023-0159 and PN-IV-P1-PCE-2023-0987.
This work was also supported by the HUN-REN Hungarian Research Network through the Supported Research Groups
Programme, HUN-REN-BME-BCE Quantum Technology Research Group (TKCS-2024/34) as well as by the Deutsche Forschungsgemeinschaft under the cluster of excellence ctd.qmat (EXC 2147, project-id 390858490).

\end{acknowledgments}

\bibliographystyle{apsrev}
\bibliography{wboson1}

\begin{thebibliography}{10}
\expandafter\ifx\csname bibnamefont\endcsname\relax
  \def\bibnamefont#1{#1}\fi
\expandafter\ifx\csname bibfnamefont\endcsname\relax
  \def\bibfnamefont#1{#1}\fi
\expandafter\ifx\csname url\endcsname\relax
  \def\url#1{\texttt{#1}}\fi
\expandafter\ifx\csname urlprefix\endcsname\relax\def\urlprefix{URL }\fi
\providecommand{\bibinfo}[2]{#2}
\providecommand{\eprint}[2][]{\url{#2}}

\bibitem{kitaev}
\bibinfo{author}{\bibfnamefont{A.}~\bibnamefont{Kitaev}},
  \emph{\bibinfo{title}{A simple model of quantum holography}},
  \bibinfo{note}{online.kitp.ucsb.edu/online/entangled15/kitaev/,
  online.kitp.ucsb.edu/online/entangled15/kitaev2/. Talks at KITP, April 7,
  2015 and May 27, 2015.}

\bibitem{kitaev17}
\bibinfo{author}{\bibfnamefont{A.}~\bibnamefont{Kitaev}} \bibnamefont{and}
  \bibinfo{author}{\bibfnamefont{S.~J.} \bibnamefont{Suh}},
  \emph{\bibinfo{title}{The soft mode in the sachdev-ye-kitaev model and its
  gravity dual}}, \bibinfo{journal}{J. High Energ. Phys.}
  \textbf{\bibinfo{volume}{2018}}, \bibinfo{pages}{183} (\bibinfo{year}{2018}).

\bibitem{Garciamata}
\bibinfo{author}{\bibfnamefont{I.}~\bibnamefont{García-Mata}},
  \bibinfo{author}{\bibfnamefont{R.~A.} \bibnamefont{Jalabert}},
  \bibnamefont{and} \bibinfo{author}{\bibfnamefont{D.~A.}
  \bibnamefont{Wisniacki}}, \emph{\bibinfo{title}{{O}ut-of-time-order
  correlations and quantum chaos}}, \bibinfo{journal}{Scholarpedia}
  \textbf{\bibinfo{volume}{18}}(\bibinfo{number}{4}), \bibinfo{pages}{55237}
  (\bibinfo{year}{2023}), \bibinfo{note}{revision \#204529}.

\bibitem{swingleprx}
\bibinfo{author}{\bibfnamefont{S.}~\bibnamefont{Xu}} \bibnamefont{and}
  \bibinfo{author}{\bibfnamefont{B.}~\bibnamefont{Swingle}},
  \emph{\bibinfo{title}{Scrambling dynamics and out-of-time-ordered correlators
  in quantum many-body systems}}, \bibinfo{journal}{PRX Quantum}
  \textbf{\bibinfo{volume}{5}}, \bibinfo{pages}{010201} (\bibinfo{year}{2024}).

\bibitem{scienceotoc}
\bibinfo{author}{\bibfnamefont{X.}~\bibnamefont{Mi}},
  \bibinfo{author}{\bibfnamefont{P.}~\bibnamefont{Roushan}},
  \bibinfo{author}{\bibfnamefont{C.}~\bibnamefont{Quintana}},
  \bibinfo{author}{\bibfnamefont{S.}~\bibnamefont{Mandrà}},
  \bibinfo{author}{\bibfnamefont{J.}~\bibnamefont{Marshall}},
  \bibinfo{author}{\bibfnamefont{C.}~\bibnamefont{Neill}},
  \bibinfo{author}{\bibfnamefont{F.}~\bibnamefont{Arute}},
  \bibinfo{author}{\bibfnamefont{K.}~\bibnamefont{Arya}},
  \bibinfo{author}{\bibfnamefont{J.}~\bibnamefont{Atalaya}},
  \bibinfo{author}{\bibfnamefont{R.}~\bibnamefont{Babbush}},
  \bibinfo{author}{\bibfnamefont{J.~C.} \bibnamefont{Bardin}},
  \bibinfo{author}{\bibfnamefont{R.}~\bibnamefont{Barends}}, \emph{et~al.},
  \emph{\bibinfo{title}{Information scrambling in quantum circuits}},
  \bibinfo{journal}{Science}
  \textbf{\bibinfo{volume}{374}}(\bibinfo{number}{6574}), \bibinfo{pages}{1479}
  (\bibinfo{year}{2021}),
  \eprint{https://www.science.org/doi/pdf/10.1126/science.abg5029}.

\bibitem{landsman}
\bibinfo{author}{\bibfnamefont{K.~A.} \bibnamefont{Landsman}},
  \bibinfo{author}{\bibfnamefont{C.}~\bibnamefont{Figgatt}},
  \bibinfo{author}{\bibfnamefont{T.}~\bibnamefont{Schuster}},
  \bibinfo{author}{\bibfnamefont{N.~M.} \bibnamefont{Linke}},
  \bibinfo{author}{\bibfnamefont{B.}~\bibnamefont{Yoshida}},
  \bibinfo{author}{\bibfnamefont{N.~Y.} \bibnamefont{Yao}}, \bibnamefont{and}
  \bibinfo{author}{\bibfnamefont{C.}~\bibnamefont{Monroe}},
  \emph{\bibinfo{title}{Verified quantum information scrambling}},
  \bibinfo{journal}{Nature} \textbf{\bibinfo{volume}{567}}, \bibinfo{pages}{61}
  (\bibinfo{year}{2019}).

\bibitem{larkin}
\bibinfo{author}{\bibfnamefont{A.~I.} \bibnamefont{Larkin}} \bibnamefont{and}
  \bibinfo{author}{\bibfnamefont{Y.~N.} \bibnamefont{Ovchinnikov}},
  \emph{\bibinfo{title}{Quasiclassical method in the theory of
  superconductivity}}, \bibinfo{journal}{Sov. Phys. JETP}
  \textbf{\bibinfo{volume}{28}}, \bibinfo{pages}{1200} (\bibinfo{year}{1969}).

\bibitem{roberts2016}
\bibinfo{author}{\bibfnamefont{D.~A.} \bibnamefont{Roberts}} \bibnamefont{and}
  \bibinfo{author}{\bibfnamefont{B.}~\bibnamefont{Swingle}},
  \emph{\bibinfo{title}{Lieb-robinson bound and the butterfly effect in quantum
  field theories}}, \bibinfo{journal}{Phys. Rev. Lett.}
  \textbf{\bibinfo{volume}{117}}, \bibinfo{pages}{091602}
  (\bibinfo{year}{2016}).

\bibitem{Stanford2016}
\bibinfo{author}{\bibfnamefont{D.}~\bibnamefont{Stanford}},
  \emph{\bibinfo{title}{Many-body chaos at weak coupling}},
  \bibinfo{journal}{Journal of High Energy Physics}
  \textbf{\bibinfo{volume}{2016}}, \bibinfo{pages}{9} (\bibinfo{year}{2016}).

\bibitem{aleiner}
\bibinfo{author}{\bibfnamefont{I.~L.} \bibnamefont{Aleiner}},
  \bibinfo{author}{\bibfnamefont{L.}~\bibnamefont{Faoro}}, \bibnamefont{and}
  \bibinfo{author}{\bibfnamefont{L.~B.} \bibnamefont{Ioffe}},
  \emph{\bibinfo{title}{Microscopic model of quantum butterfly effect:
  Out-of-time-order correlators and traveling combustion waves}},
  \bibinfo{journal}{Annals of Physics} \textbf{\bibinfo{volume}{375}},
  \bibinfo{pages}{378} (\bibinfo{year}{2016}).

\bibitem{garttner}
\bibinfo{author}{\bibfnamefont{M.}~\bibnamefont{G{\"a}rttner}},
  \bibinfo{author}{\bibfnamefont{J.~G.} \bibnamefont{Bohnet}},
  \bibinfo{author}{\bibfnamefont{A.}~\bibnamefont{{Safavi-Naini}}},
  \bibinfo{author}{\bibfnamefont{M.~L.} \bibnamefont{Wall}},
  \bibinfo{author}{\bibfnamefont{J.~J.} \bibnamefont{Bollinger}},
  \bibnamefont{and} \bibinfo{author}{\bibfnamefont{A.~M.} \bibnamefont{Rey}},
  \emph{\bibinfo{title}{Measuring out-of-time-order correlations and multiple
  quantum spectra in a trapped ion quantum magnet}}, \bibinfo{journal}{Nature
  Phys} \textbf{\bibinfo{volume}{13}}, \bibinfo{pages}{781}
  (\bibinfo{year}{2017}).

\bibitem{junli}
\bibinfo{author}{\bibfnamefont{J.}~\bibnamefont{Li}},
  \bibinfo{author}{\bibfnamefont{R.}~\bibnamefont{Fan}},
  \bibinfo{author}{\bibfnamefont{H.}~\bibnamefont{Wang}},
  \bibinfo{author}{\bibfnamefont{B.}~\bibnamefont{Ye}},
  \bibinfo{author}{\bibfnamefont{B.}~\bibnamefont{Zeng}},
  \bibinfo{author}{\bibfnamefont{H.}~\bibnamefont{Zhai}},
  \bibinfo{author}{\bibfnamefont{X.}~\bibnamefont{Peng}}, \bibnamefont{and}
  \bibinfo{author}{\bibfnamefont{J.}~\bibnamefont{Du}},
  \emph{\bibinfo{title}{Measuring out-of-time-order correlators on a nuclear
  magnetic resonance quantum simulator}}, \bibinfo{journal}{Phys. Rev. X}
  \textbf{\bibinfo{volume}{7}}, \bibinfo{pages}{031011} (\bibinfo{year}{2017}).

\bibitem{maldacenaprd}
\bibinfo{author}{\bibfnamefont{J.}~\bibnamefont{Maldacena}} \bibnamefont{and}
  \bibinfo{author}{\bibfnamefont{D.}~\bibnamefont{Stanford}},
  \emph{\bibinfo{title}{Remarks on the sachdev-ye-kitaev model}},
  \bibinfo{journal}{Phys. Rev. D} \textbf{\bibinfo{volume}{94}},
  \bibinfo{pages}{106002} (\bibinfo{year}{2016}).

\bibitem{zhugrover}
\bibinfo{author}{\bibfnamefont{G.}~\bibnamefont{Zhu}},
  \bibinfo{author}{\bibfnamefont{M.}~\bibnamefont{Hafezi}}, \bibnamefont{and}
  \bibinfo{author}{\bibfnamefont{T.}~\bibnamefont{Grover}},
  \emph{\bibinfo{title}{Measurement of many-body chaos using a quantum clock}},
  \bibinfo{journal}{Phys. Rev. A} \textbf{\bibinfo{volume}{94}},
  \bibinfo{pages}{062329} (\bibinfo{year}{2016}).

\bibitem{prxkeyserlingk}
\bibinfo{author}{\bibfnamefont{C.~W.} \bibnamefont{von Keyserlingk}},
  \bibinfo{author}{\bibfnamefont{T.}~\bibnamefont{Rakovszky}},
  \bibinfo{author}{\bibfnamefont{F.}~\bibnamefont{Pollmann}}, \bibnamefont{and}
  \bibinfo{author}{\bibfnamefont{S.~L.} \bibnamefont{Sondhi}},
  \emph{\bibinfo{title}{Operator hydrodynamics, otocs, and entanglement growth
  in systems without conservation laws}}, \bibinfo{journal}{Phys. Rev. X}
  \textbf{\bibinfo{volume}{8}}, \bibinfo{pages}{021013} (\bibinfo{year}{2018}).

\bibitem{maldacena2016}
\bibinfo{author}{\bibfnamefont{J.}~\bibnamefont{Maldacena}},
  \bibinfo{author}{\bibfnamefont{S.~H.} \bibnamefont{Shenker}},
  \bibnamefont{and} \bibinfo{author}{\bibfnamefont{D.}~\bibnamefont{Stanford}},
  \emph{\bibinfo{title}{A bound on chaos}}, \bibinfo{journal}{Journal of High
  Energy Physics} \textbf{\bibinfo{volume}{2016}}, \bibinfo{pages}{106}
  (\bibinfo{year}{2016}).

\bibitem{bohrdt}
\bibinfo{author}{\bibfnamefont{A.}~\bibnamefont{Bohrdt}},
  \bibinfo{author}{\bibfnamefont{C.~B.} \bibnamefont{Mendl}},
  \bibinfo{author}{\bibfnamefont{M.}~\bibnamefont{Endres}}, \bibnamefont{and}
  \bibinfo{author}{\bibfnamefont{M.}~\bibnamefont{Knap}},
  \emph{\bibinfo{title}{Scrambling and thermalization in a diffusive quantum
  many-body system}}, \bibinfo{journal}{New Journal of Physics}
  \textbf{\bibinfo{volume}{19}}(\bibinfo{number}{6}), \bibinfo{pages}{063001}
  (\bibinfo{year}{2017}).

\bibitem{roberts}
\bibinfo{author}{\bibfnamefont{D.~A.} \bibnamefont{Roberts}} \bibnamefont{and}
  \bibinfo{author}{\bibfnamefont{D.}~\bibnamefont{Stanford}},
  \emph{\bibinfo{title}{Diagnosing chaos using four-point functions in
  two-dimensional conformal field theory}}, \bibinfo{journal}{Phys. Rev. Lett.}
  \textbf{\bibinfo{volume}{115}}, \bibinfo{pages}{131603}
  (\bibinfo{year}{2015}).

\bibitem{roberts2017}
\bibinfo{author}{\bibfnamefont{D.~A.} \bibnamefont{Roberts}} \bibnamefont{and}
  \bibinfo{author}{\bibfnamefont{B.}~\bibnamefont{Yoshida}},
  \emph{\bibinfo{title}{Chaos and complexity by design}}, \bibinfo{journal}{J.
  High Energ. Phys.} \textbf{\bibinfo{volume}{2017}}, \bibinfo{pages}{121}
  (\bibinfo{year}{2017}).

\bibitem{googleotoc}
\bibinfo{author}{\bibnamefont{{Google Quantum AI and Collaborators}}},
  \emph{\bibinfo{title}{Observation of constructive interference at the edge of
  quantum ergodicity}}, \bibinfo{journal}{Nature}
  \textbf{\bibinfo{volume}{646}}, \bibinfo{pages}{825} (\bibinfo{year}{2025}).

\bibitem{iaconis}
\bibinfo{author}{\bibfnamefont{J.}~\bibnamefont{Iaconis}},
  \emph{\bibinfo{title}{Quantum state complexity in computationally tractable
  quantum circuits}}, \bibinfo{journal}{PRX Quantum}
  \textbf{\bibinfo{volume}{2}}, \bibinfo{pages}{010329} (\bibinfo{year}{2021}).

\bibitem{fritzsch}
\bibinfo{author}{\bibfnamefont{F.}~\bibnamefont{Fritzsch}},
  \bibinfo{author}{\bibfnamefont{G.~O.} \bibnamefont{Alves}},
  \bibinfo{author}{\bibfnamefont{M.~A.} \bibnamefont{Rampp}}, \bibnamefont{and}
  \bibinfo{author}{\bibfnamefont{P.~W.} \bibnamefont{Claeys}},
  \emph{\bibinfo{title}{Free cumulants and full eigenstate thermalization from
  boundary scrambling}}, \bibinfo{note}{{a}rXiv:2509.08060}.

\bibitem{giamarchi}
\bibinfo{author}{\bibfnamefont{T.}~\bibnamefont{Giamarchi}},
  \emph{\bibinfo{title}{Quantum Physics in One Dimension}}
  (\bibinfo{publisher}{Oxford University Press}, \bibinfo{address}{Oxford},
  \bibinfo{year}{2004}).

\bibitem{nersesyan}
\bibinfo{author}{\bibfnamefont{A.~O.} \bibnamefont{Gogolin}},
  \bibinfo{author}{\bibfnamefont{A.~A.} \bibnamefont{Nersesyan}},
  \bibnamefont{and} \bibinfo{author}{\bibfnamefont{A.~M.}
  \bibnamefont{Tsvelik}}, \emph{\bibinfo{title}{Bosonization and Strongly
  Correlated Systems}} (\bibinfo{publisher}{Cambridge University Press},
  \bibinfo{address}{Cambridge}, \bibinfo{year}{1998}).

\bibitem{cazalillaboson}
\bibinfo{author}{\bibfnamefont{M.~A.} \bibnamefont{Cazalilla}},
  \emph{\bibinfo{title}{Bosonizing one-dimensional cold atomic gases}},
  \bibinfo{journal}{J. Phys. B: At. Mol. Opt. Phys.}
  \textbf{\bibinfo{volume}{37}}, \bibinfo{pages}{S1} (\bibinfo{year}{2004}).

\bibitem{nielsen}
\bibinfo{author}{\bibfnamefont{M.}~\bibnamefont{Nielsen}} \bibnamefont{and}
  \bibinfo{author}{\bibfnamefont{I.}~\bibnamefont{Chuang}},
  \emph{\bibinfo{title}{Quantum Computation and Quantum Information}}
  (\bibinfo{publisher}{Cambridge University Press},
  \bibinfo{address}{Cambridge}, \bibinfo{year}{2000}).

\bibitem{favaprx}
\bibinfo{author}{\bibfnamefont{M.}~\bibnamefont{Fava}},
  \bibinfo{author}{\bibfnamefont{J.}~\bibnamefont{Kurchan}}, \bibnamefont{and}
  \bibinfo{author}{\bibfnamefont{S.}~\bibnamefont{Pappalardi}},
  \emph{\bibinfo{title}{Designs via free probability}}, \bibinfo{journal}{Phys.
  Rev. X} \textbf{\bibinfo{volume}{15}}, \bibinfo{pages}{011031}
  (\bibinfo{year}{2025}).

\bibitem{delft}
\bibinfo{author}{\bibfnamefont{J.}~\bibnamefont{von Delft}} \bibnamefont{and}
  \bibinfo{author}{\bibfnamefont{H.}~\bibnamefont{Schoeller}},
  \emph{\bibinfo{title}{Bosonization for beginners - refermionization for
  experts}}, \bibinfo{journal}{Ann. Phys. (Leipzig)}
  \textbf{\bibinfo{volume}{7}}, \bibinfo{pages}{225} (\bibinfo{year}{1998}).

\bibitem{luther}
\bibinfo{author}{\bibfnamefont{A.}~\bibnamefont{Luther}} \bibnamefont{and}
  \bibinfo{author}{\bibfnamefont{I.}~\bibnamefont{Peschel}},
  \emph{\bibinfo{title}{Calculation of critical exponents in two dimensions
  from quantum field theory in one dimension}}, \bibinfo{journal}{Phys. Rev. B}
  \textbf{\bibinfo{volume}{12}}, \bibinfo{pages}{3908} (\bibinfo{year}{1975}).

\bibitem{iucci2kf}
\bibinfo{author}{\bibfnamefont{A.}~\bibnamefont{Iucci}},
  \bibinfo{author}{\bibfnamefont{G.~A.} \bibnamefont{Fiete}}, \bibnamefont{and}
  \bibinfo{author}{\bibfnamefont{T.}~\bibnamefont{Giamarchi}},
  \emph{\bibinfo{title}{Fourier transform of the $2{k}_{F}$ luttinger liquid
  density correlation function with different spin and charge velocities}},
  \bibinfo{journal}{Phys. Rev. B} \textbf{\bibinfo{volume}{75}},
  \bibinfo{pages}{205116} (\bibinfo{year}{2007}).

\bibitem{Note1}
\bibinfo{note}{We use $\cos (6c)=4\cos ^3(2c)-3\cos (2c)$ to rewrite all
  harmonics but the highest (i.e. $10c$) in terms of $\cos (2c)$ when comparing
  to numerics.}

\bibitem{harper}
\bibinfo{author}{\bibfnamefont{P.~G.} \bibnamefont{Harper}},
  \emph{\bibinfo{title}{Single band motion of conduction electrons in a uniform
  magnetic field}}, \bibinfo{journal}{Proceedings of the Physical Society.
  Section A} \textbf{\bibinfo{volume}{68}}(\bibinfo{number}{10}),
  \bibinfo{pages}{874} (\bibinfo{year}{1955}).

\bibitem{aubryandre}
\bibinfo{author}{\bibfnamefont{A.}~\bibnamefont{Aubry}} \bibnamefont{and}
  \bibinfo{author}{\bibfnamefont{G.}~\bibnamefont{Andre}},
  \emph{\bibinfo{title}{Analyticity breaking and anderson localization in
  incommensurate lattices}}, \bibinfo{journal}{Ann. Israel. Phys. Soc.}
  \textbf{\bibinfo{volume}{3}}, \bibinfo{pages}{133} (\bibinfo{year}{1980}).

\bibitem{quillen}
\bibinfo{author}{\bibfnamefont{A.~C.} \bibnamefont{Quillen}},
  \bibinfo{author}{\bibfnamefont{N.}~\bibnamefont{Skerrett}},
  \bibinfo{author}{\bibfnamefont{D.~R.} \bibnamefont{Sowinski}},
  \bibnamefont{and} \bibinfo{author}{\bibfnamefont{A.~S.}
  \bibnamefont{Miakhel}}, \emph{\bibinfo{title}{Notions of adiabatic drift in
  the quantized harper model}}, \bibinfo{journal}{Phys. Rev. A}
  \textbf{\bibinfo{volume}{112}}, \bibinfo{pages}{042226}
  (\bibinfo{year}{2025}).

\bibitem{moghaddan}
\bibinfo{author}{\bibfnamefont{A.~G.} \bibnamefont{Moghaddam}},
  \bibinfo{author}{\bibfnamefont{V.}~\bibnamefont{K\"onye}},
  \bibinfo{author}{\bibfnamefont{L.}~\bibnamefont{Mertens}},
  \bibinfo{author}{\bibfnamefont{J.}~\bibnamefont{van Wezel}},
  \bibnamefont{and} \bibinfo{author}{\bibfnamefont{J.}~\bibnamefont{van~den
  Brink}}, \emph{\bibinfo{title}{Synthetic horizons in an atomic chain:
  Horizon-induced effects and connections to quantum hall systems}},
  \bibinfo{journal}{Phys. Rev. B} \textbf{\bibinfo{volume}{113}},
  \bibinfo{pages}{195430} (\bibinfo{year}{2026}).

\bibitem{xuortiz}
\bibinfo{author}{\bibfnamefont{Q.-R.} \bibnamefont{Xu}},
  \bibinfo{author}{\bibfnamefont{E.}~\bibnamefont{Cobanera}}, \bibnamefont{and}
  \bibinfo{author}{\bibfnamefont{G.}~\bibnamefont{Ortiz}},
  \emph{\bibinfo{title}{Bloch and bethe ans\"atze for the harper model: A
  butterfly with a boundary}}, \bibinfo{journal}{Phys. Rev. B}
  \textbf{\bibinfo{volume}{104}}, \bibinfo{pages}{165140}
  (\bibinfo{year}{2021}).

\bibitem{korsch}
\bibinfo{author}{\bibfnamefont{H.~J.} \bibnamefont{Korsch}} \bibnamefont{and}
  \bibinfo{author}{\bibfnamefont{K.}~\bibnamefont{Rapedius}},
  \emph{\bibinfo{title}{Computations in quantum mechanics made easy}},
  \bibinfo{journal}{European Journal of Physics}
  \textbf{\bibinfo{volume}{37}}(\bibinfo{number}{5}), \bibinfo{pages}{055410}
  (\bibinfo{year}{2016}).

\bibitem{lukyanov}
\bibinfo{author}{\bibfnamefont{S.}~\bibnamefont{Lukyanov}},
  \emph{\bibinfo{title}{Correlation amplitude for the \textit{XXZ} spin chain
  in the disordered regime}}, \bibinfo{journal}{Phys. Rev. B}
  \textbf{\bibinfo{volume}{59}}, \bibinfo{pages}{11163} (\bibinfo{year}{1999}).

\bibitem{doraotoc}
\bibinfo{author}{\bibfnamefont{B.}~\bibnamefont{D\'ora}} \bibnamefont{and}
  \bibinfo{author}{\bibfnamefont{R.}~\bibnamefont{Moessner}},
  \emph{\bibinfo{title}{Out-of-time-ordered density correlators in luttinger
  liquids}}, \bibinfo{journal}{Phys. Rev. Lett.}
  \textbf{\bibinfo{volume}{119}}, \bibinfo{pages}{026802}
  (\bibinfo{year}{2017}).

\bibitem{motrunich}
\bibinfo{author}{\bibfnamefont{C.-J.} \bibnamefont{Lin}} \bibnamefont{and}
  \bibinfo{author}{\bibfnamefont{O.~I.} \bibnamefont{Motrunich}},
  \emph{\bibinfo{title}{Out-of-time-ordered correlators in short-range and
  long-range hard-core boson models and in the luttinger-liquid model}},
  \bibinfo{journal}{Phys. Rev. B} \textbf{\bibinfo{volume}{98}},
  \bibinfo{pages}{134305} (\bibinfo{year}{2018}).

\bibitem{Gisti2025}
\bibinfo{author}{\bibfnamefont{M.}~\bibnamefont{Gisti}},
  \bibinfo{author}{\bibfnamefont{D.~J.} \bibnamefont{Luitz}}, \bibnamefont{and}
  \bibinfo{author}{\bibfnamefont{M.}~\bibnamefont{Debertolis}},
  \emph{\bibinfo{title}{Symmetry resolved out-of-time-order correlators of
  {H}eisenberg spin chains using projected matrix product operators}},
  \bibinfo{journal}{{Quantum}} \textbf{\bibinfo{volume}{9}},
  \bibinfo{pages}{1871} (\bibinfo{year}{2025}), ISSN \bibinfo{issn}{2521-327X}.

\bibitem{hosur}
\bibinfo{author}{\bibfnamefont{P.}~\bibnamefont{Hosur}},
  \bibinfo{author}{\bibfnamefont{X.-L.} \bibnamefont{Qi}},
  \bibinfo{author}{\bibfnamefont{D.~A.} \bibnamefont{Roberts}},
  \bibnamefont{and} \bibinfo{author}{\bibfnamefont{B.}~\bibnamefont{Yoshida}},
  \emph{\bibinfo{title}{Chaos in quantum channels}}, \bibinfo{journal}{J. High
  Energ. Phys} \textbf{\bibinfo{volume}{2016}}, \bibinfo{pages}{4}
  (\bibinfo{year}{2016}).

\end{thebibliography}

\end{document}